# Structural Inversion Asymmetry in Epitaxial Ultrathin Films of Bi (111)/InSb (111)B

Hadass S. Inbar[1*&], Muhammad Zubair[2,3&], Jason T. Dong[1], Aaron N Engel[1], Connor P. Dempsey[4], Yu Hao Chang[1], Shinichi Nishihaya[1†], Shoaib Khalid[5], Alexei V. Fedorov[6], Anderson Janotti[3], Chris J. Palmstrøm[1,4*]

[1] *Materials Department, UC Santa Barbara, CA 93106*

[2] *Department of Physics and Astronomy, University of Delaware, Newark, DE 19716, USA*

[3] *Department of Materials Science and Engineering, University of Delaware, Newark, DE 19716, USA*

[4] *Department of Electrical & Computer Engineering, UC Santa Barbara, CA 93106*

[5] *Princeton Plasma Physics Laboratory, Princeton, NJ 08540, USA*

[6] *Advanced Light Source, Lawrence Berkeley National Laboratory, Berkeley, CA 94720, USA*

Bismuth (Bi) films hold potential for spintronic devices due to strong spin-orbit coupling. Understanding the growth, surface states, and interactions with the substrate is key to their functionalization. Large-area high-quality (111) Bi ultrathin films were grown on InSb (111)B substrates by molecular beam epitaxy (MBE). Strong film-substrate interactions epitaxially stabilize the (111) orientation and lead to nonequivalent interface potentials. Analysis of angle-resolved photoemission spectroscopy (ARPES) measurements, employed to characterize the evolution of the surface states with film thickness, indicate a crossing at the $\overline{M}$ point, suggesting a topologically trivial phase in the thin film. The results show the presence of interfacial bonds to the substrate breaks inversion symmetry, preventing the semimetal-to-semiconductor transition predicted for freestanding bismuth layers, highlighting the importance of controlled functionalization and surface passivation of two-dimensional materials.

## 1. INTRODUCTION

Scientists have studied strain and quantum size effects in bismuth (Bi) films for decades, which provide a rich platform for tuning topological order [1], semimetal to semiconducting transitions [2], and quantum-well states (QWS) [3]. The low carrier density, long mean free path, large spin-orbit coupling, and spin-polarized surface states [4] make Bi films a promising system for future spintronics applications [5]. Group-V semimetals crystalize in the rhombohedral A7 structure in bulk, but their two-dimensional (2D) variants can stabilize in additional phases [6]. Group-V 2D films have attracted interest in classical electronic and optoelectronic device applications due to their high carrier mobility and potential bandgap tunability [7]. In the field of topological materials, there is an ongoing effort to classify the $\mathbb{Z}_2$ invariant of Bi experimentally [8–11], which also proves challenging to calculate computationally [12]. 1D helical modes were observed along the Bi (111) type A step edges[13,14], which are a key component in one proposed platform to construct Majorana zero modes [15].

The synthesis of large-area single-domain ultrathin (<6 bilayers, BL) buckled Bi $(111)_r$ on conventional semiconducting substrates has remained a challenge, with only planar bismuthene wetting layers on SiC [16] and GaAs [17] reported thus far. On weakly interacting substrates, such as highly oriented pyrolytic graphite, Bi nucleates typically in the black phosphorus (BP)-like phase [18,19], and transforms upon further deposition to the rhombohedral $(111)_r$ oriented phase. On Si (111) substrates, weak film-substrate van der Waals (vdW) interactions [20,21] also lead to the nucleation of nearly freestanding Bi layers, starting at the BP phase and transforming to a $(111)_r$ orientation only after a 6 BL film coalesces [6]. Ultrathin Bi (111) films were nucleated on the topological insulator (TI) substrate $Bi_2Te_3$ [22–25], where in-plane contraction [26,27] is suggested to stabilize a TI phase. However, compressive strain and band hybridization with the $Bi_2Te_3$ substrate, along with unclear correspondence between the experimental data and the calculations at the ultrathin limit [23], make it difficult to study the topological classification of Bi and the semimetal to semiconducting transition predicted for ultrathin Bi [28].

Bulk Bi is a semimetal [Fig. 1, (a) and (b)] with a valence-band maximum at the time-reversal invariant momentum (TRIM) $T$ point (projecting to $\overline{\Gamma}$ for the (111) surface) and a conduction-band minimum at the $L$ TRIM point (projecting to $\overline{M}$). The small direct bandgap at $L$ is only a few meV and determines whether Bi is a strong TI (an inverted bandgap at $L$) or a higher-order TI (HOTI, no band inversion at $L$). Bi is predicted to lie at the border of a topological phase transition between the HOTI and TI phases [9]. Yet, despite the challenge of estimating the gap size at $L$, most density functional theory (DFT) calculations predict a trivial band order, with quasiparticle self-consistent $GW$ calculations yielding a gap of 13 meV compared to 86 meV in conventional DFT [1]. The indirect $T$-$L$ gap and the

& These authors contributed equally

Present Address: † Department of Physics, Tokyo Institute of Technology, Tokyo, 152-8551, Japan

Corresponding authors: * Email: hadass@ucsb.edu (H.S.I.), cjpalm@ucsb.edu (C.J.P.)

bandgap inversion at $L$ depend on electron density [29], biaxial and shear strain [1,9,23], and bulk alloying in $Bi_{1-x}Sb_x$ [30].

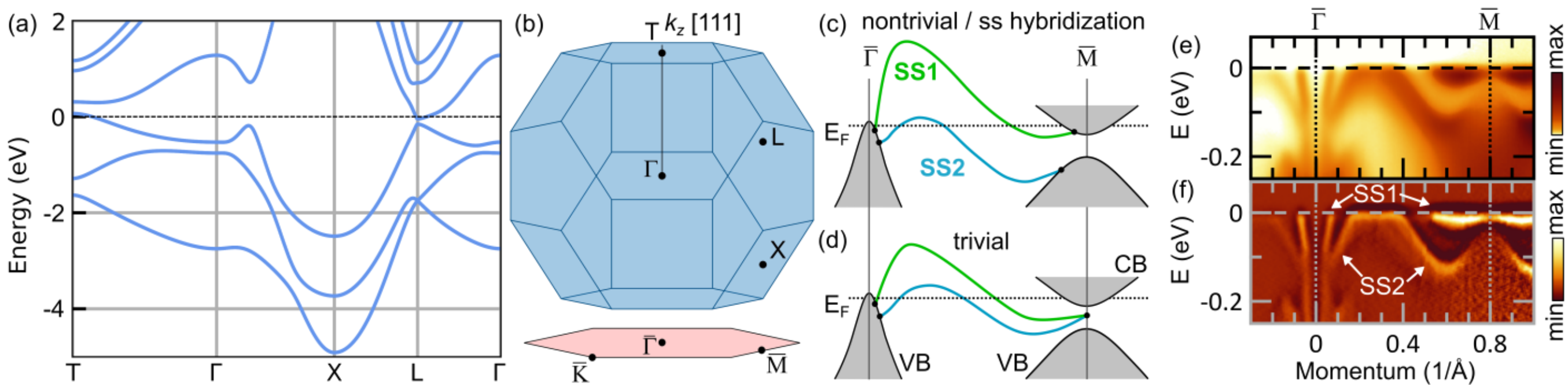

Fig. 1. (a) Calculated band structure of bulk Bi showing the hole and electron pockets at TRIM points $T$ and $L$, respectively. (b) The bulk Brillouin zone and TRIM points projected onto the (111) surface Brillouin zone. (c), (d) Schematic drawings of two possible surface-state (SS1 and SS2) dispersions and projected valence and conduction bands (VB, CB) along $\bar{\Gamma} - \bar{M}$; $E_F$ is the Fermi energy. (c) A surface state gap could indicate either a $\mathbb{Z}_2$ topologically nontrivial band structure for a semi-infinite crystal or interactions between surface states at the top and bottom surfaces of a film. (d) Surface state degeneracy at $\bar{M}$, indicating a $\mathbb{Z}_2$ trivial band structure. (e), (f) ARPES $E$-$k$ dispersion of the surface states at $h\nu$ = 37.5 eV for a 200 BL Bi (111) /InSb (111)B. (e) Raw data and (f) curvature plot [31] of the raw data enhancing the dispersive features.

Angle-resolved photoemission spectroscopy (ARPES) measurements of Bi films [8,32,33] have nonetheless shown surface states gapped at $\bar{M}$ which in the past was attributed to a $\mathbb{Z}_2$ nontrivial band topology [Fig. 1(c)] based on the surface state band-counting criteria [30,33]. ARPES observations of surface state connectivity can be used to classify topological phases in quantum materials [34,35], but they are complicated by spectral weight variations and surface-bulk hybridization [36], and their energy and momentum resolution may be limited [37]. Moreover, in thinner films, the spatial overlap between the wave functions of surface states at the top and bottom interfaces could also open up a hybridization gap [12]. In Bi (111), the surface states near $\bar{M}$ penetrate into the bulk of the film by 100s of BLs [12,38,39]. Crosstalk between the surface states forms a hybridization gap even in 200 BL thick films [8], resulting in an inconclusive assignment of the topological phase [Fig. 1(c)] [12].

Recent *ab initio* calculations [40] have predicted that in films with structural inversion asymmetry (SIA), one could more easily distinguish between the $\mathbb{Z}_2$ topologically trivial and nontrivial phases due to the emergence of degenerate surface states observed only from one surface, as schematically portrayed in Fig. 1(d). SIA in Bi thin films was modeled in past publications using surface potentials/functionalization [40–42] or inter/intra-BL expansion and structural changes [12,42]. However, not all interface perturbations are significant enough to prevent the neighboring surface states from hybridizing. For example, small potential differences in weakly interacting substrates such as $Bi/Bi_2Te_3$ and Bi/Si still result in surface state hybridization. Thus, to study the true topological nature of Bi, a substrate/overlayer with strong bonding to Bi films is necessary to produce SIA with an energy difference large enough to prevent surface states from hybridizing and facilitate the topological phase assignment of Bi. In addition to experimental efforts to tune the external electric field across the Bi film thickness using substrate chemistry and bias, further theoretical modeling of Bi thin films is needed. This modeling should focus on identifying substrate-induced voltage phase transitions between topologically trivial and nontrivial phases [43,44].

In this work, we utilize a semiconducting substrate, InSb (111)B, with strong film-substrate interaction, resulting in two non-equivalent interfaces (substrate-Bi and Bi-vacuum). Unlike previously explored substrates, these strong interactions allow us to stabilize ultrathin large-area (111) Bi films down to 1 BL in thickness while avoiding neighboring surface state hybridization. Our first-principles electronic structure calculations confirm the strong effect of the substrate on the surface states of the Bi thin films. The results indicate a complete decoupling of the top and bottom surfaces reflected in the degeneracy of the surface states at the $\bar{M}$ TRIM point, indicative of a topologically trivial phase. However, conclusive determination of the topological phase of Bi/InSb requires higher resolution of energy-momentum mapping of the Fermi contour around $\bar{M}$. Moreover, additional work is needed to identify any presence of a critical SIA potential driving the films between a trivial and a potentially nontrivial phase.

## 2. RESULTS AND DISCUSSION

The Bi films are grown via molecular beam epitaxy (see Section S1 [45]) on InSb (111)B wafers and are studied with scanning tunneling microscopy (STM), reflection high energy diffraction (RHEED), and ARPES. Fig. 2 presents the nucleation of the Bi films grown on InSb (111)B. Fig. 2(a) and (b) show the epitaxial relationship between the zincblende InSb (111)B surface and the Bi film with a rhombohedral crystal structure. The $(111)_r$ rhombohedral notation of Bi can also be simplified using the $(0001)_{hex}$ quasi-hexagonal unit cell in Fig. 2(a), where 3 BL (BL=3.95 Å) define the hexagonal out-of-plane lattice parameter $c_0$= 11.862 Å, with an in-plane lattice parameter $a_0$= 4.546 Å [46]. Each Bi BL has a buckled structure separated by a vdW-like gap, with intermediate coupling strength between the bilayers [47]. The mismatch between the Bi bulk lattice constant and InSb <110> atomic spacing is relatively small, with only 0.8% biaxial tensile stress applied to the Bi film. RHEED patterns in Fig. 2(c) confirm the nucleation of smooth (1×1) unreconstructed Bi film surface on a well-ordered (3×3) InSb (111)B surface reconstruction that is atomically smooth over a large area [Fig. 2(d), Fig. S1(a) [45]].

STM images in Fig. 2(d)-(g) show the evolution of the ultrathin film morphology with Bi film thickness. Following the deposition of 1 BL, a fractal-like structure emerges [Fig. 2(e), Fig. S1(b) [45]] which was recently reported for the same film-substrate system [48]. For 2-3 BL films, the fractal pattern transforms to nearly continuous coverage of Bi (111) with 1-2 BL steps formed by adatom/vacancy clusters [2], which in turn could lead to band broadening in the ARPES spectra. Future work optimizing the film roughness through a more extensive synthesis study could help reduce the surface roughness down to the atomic limit, by mapping ideal nucleation temperatures and post-growth annealing conditions for each film thickness.

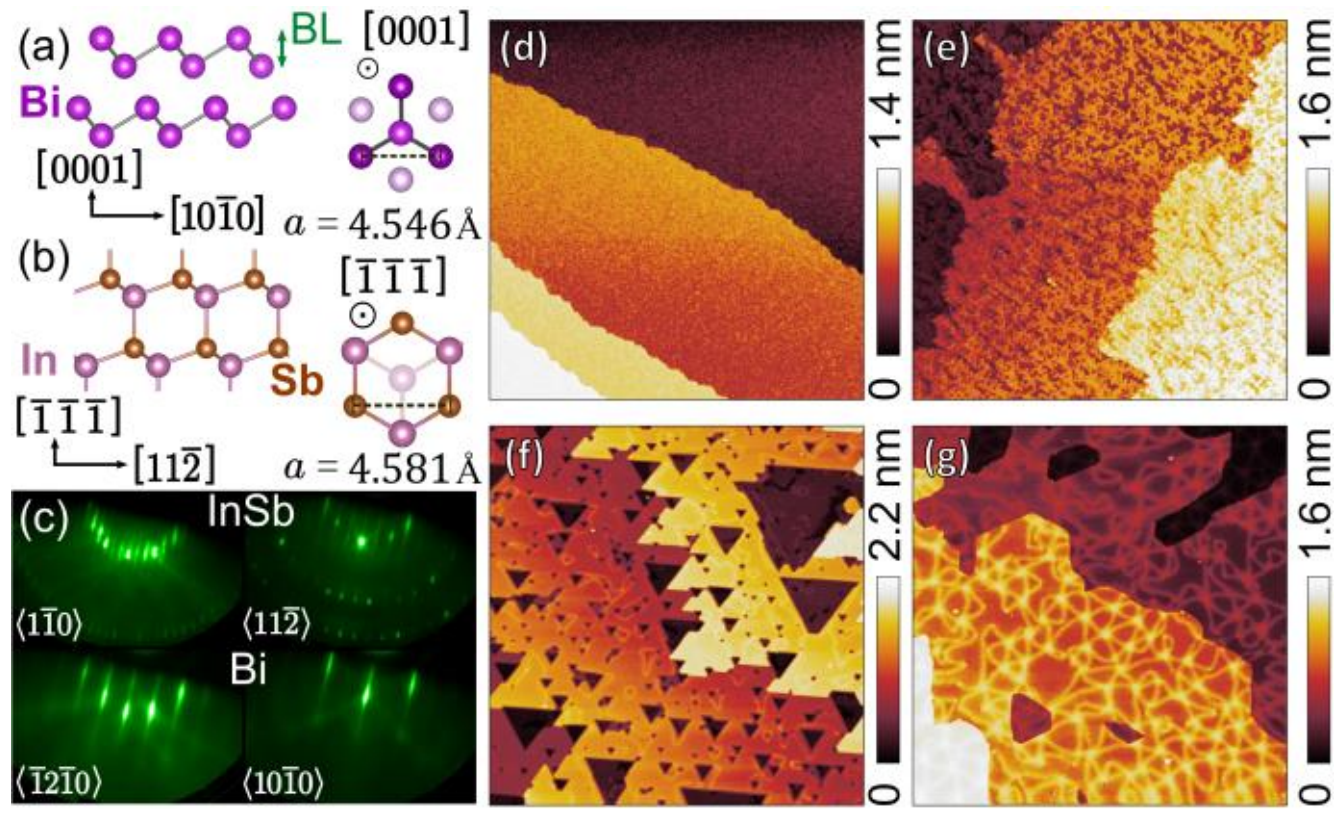

Fig. 2. Side and top-view models of (a) the Bi $(0001)_{hex}$ surface illustrating intralayer covalent-like bonds and interlayer vdW-like stacking and (b) InSb (111)B unreconstructed surface. (c) RHEED patterns and epitaxial alignment of the (3×3) InSb (111)B substrate and (1×1) Bi $(0001)_{hex}$ surface. (d)-(g) STM images of InSb and ultrathin Bi films 400×400 nm$^2$. (d) InSb (111)B substrate exhibiting atomically flat terraces (bias voltage, $V_b$: 1.2 V). (e) 1 BL Bi film ($V_b$: 3 V). (f) 2.6 BL Bi film ($V_b$: 3 V). (g) 5.4 BL thick film, showing progressive film relaxation ($V_b$: 3 V).

In Fig. 2(f), we observe only a single domain orientation in our STM images, unlike the common rotational domains seen thus far for Bi films [21]. Coherent strain is maintained only up to 2 BL [49], after which the films partially relaxes and form a “ripplocation” /soliton network, a unique strain relief mechanism in vdW materials [50]. The relaxed soliton network is evident in Fig. 2(g) and Fig. S1(c) [45] for a 5.4 BL film and was observed in Bi films only when grown on InSb [48,49]. The average terrace area observed in the 5.4 BL thick film is comparable to the steps seen for the InSb (111)B surface in Fig. 2(d), and results from substrate vicinality. This indicates an early onset of full coverage of Bi on the InSb substrate (in comparison with the much rougher nucleation on Ge/Si [6]), and overall a smooth surface of the Bi films at the ultrathin limit.

ARPES measurements of the surface states and QWS in the Bi (111)/InSb (111)B films were conducted at the Advanced Light Source at 11 K (see Section S1 [45] for details). The photon energies used in this work, $h\nu$ = 37.5 and 20 eV, should correspond to being close to the bulk $\Gamma$ and $T$ points, respectively, assuming an inner potential of 6-10 eV. Fig. 3(a) shows the Fermi surface of a 5.4 BL Bi film, presenting three-fold symmetry as the surface states disperse from the valence band at $\bar{\Gamma}$ towards $\bar{M}$, indicating a single epitaxial domain orientation. In Fig. 3(b)-(e), we follow the evolution of surface states (SS1 and SS2) and QWS as a function of film thickness near $\bar{M}$, with wide energy range scans provided in Fig. S2 [45]. An increasing degree of broadening in the surface state spectra is observed at the ultrathin limit (< 6 BL), that could be attributed to a rougher surface, either due to corrugations from terrace steps, or out-of-plane buckling of the thin Bi film due to the soliton relaxation [49]. We do not observe in ARPES any of the InSb valence band dispersions despite the ultrathin thickness of the Bi films (< 6 BL), as confirmed by examining a reference InSb (111)B surface shown in Fig. S3 [45].

Contrary to earlier reports for Bi (111) grown on other substrates, such as Si(111) [3], Ge (111) [8], or $Bi_2Te_3$ (111) [23,25], our data in Fig. 3(b) and (c) suggest a surface state band degeneracy at $\bar{M}$, consistent with the trivial surface state assignment of Bi [Fig. 1(d)] and indicative of the surface state bands avoiding crosstalk between the top and bottom surfaces. For Bi films thinner than 10BL, the surface state hybridization gap at $\bar{M}$ is expected to be at least a few 100s of meV [28]; therefore, to observe the gapped surface states, the photon energies used in our work would be sufficient. However, for the thicker films, such as the 200BL film, additional studies measuring the same structure with a UV laser source would provide an improved energy-momentum resolution needed to resolve the surface state crossing and rule out very small gaps for hybridized surface states [10]. The lack of anticipated gapped surface states, characteristic of a strong TI phase, suggests the possibility of Bi (111)/InSb(111)B films residing in the HOTI phase. In Fig. 3 and S1 we do not detect the 1D edge state modes predicted for the HOTI phase [13] (For a Bi (111) film, 1D states would serpentine along $\langle\bar{1}2\bar{1}0\rangle$ edges ($\bar{\Gamma}-\bar{K}$ ) [14,51]), likely due to their relatively weak spectral weight [51].

To elucidate the origin of the avoided surface state hybridization from the top and bottom surface of the Bi layer, we perform DFT calculations for two possible Bi film structures with thicknesses varying from 1-6 BLs: (i) an inversion-symmetric freestanding Bi slab [Fig. 4(a)] and (ii) a Bi/InSb stack [Fig. 4(b)]. See Section S1 and Section S2 for details [45]. Our band structure calculations for the freestanding layers in Fig. 4(c) are consistent with earlier studies [28,52], predicting a semimetal to semiconductor transition for 1-2 BL thick films and surface states gapped at $\bar{M}$ due to the crosstalk between the top and bottom surfaces. In contrast, the Bi/InSb stack in Fig. 4(d) shows a surface state degeneracy at $\bar{M}$ and an increasing separation in the surface state Fermi wavevectors with film thickness. These results for the Bi/InSb (111)B structure calculations agree with our experimental observations in Fig. 3(b)-(e) and Fig. S2 [45].

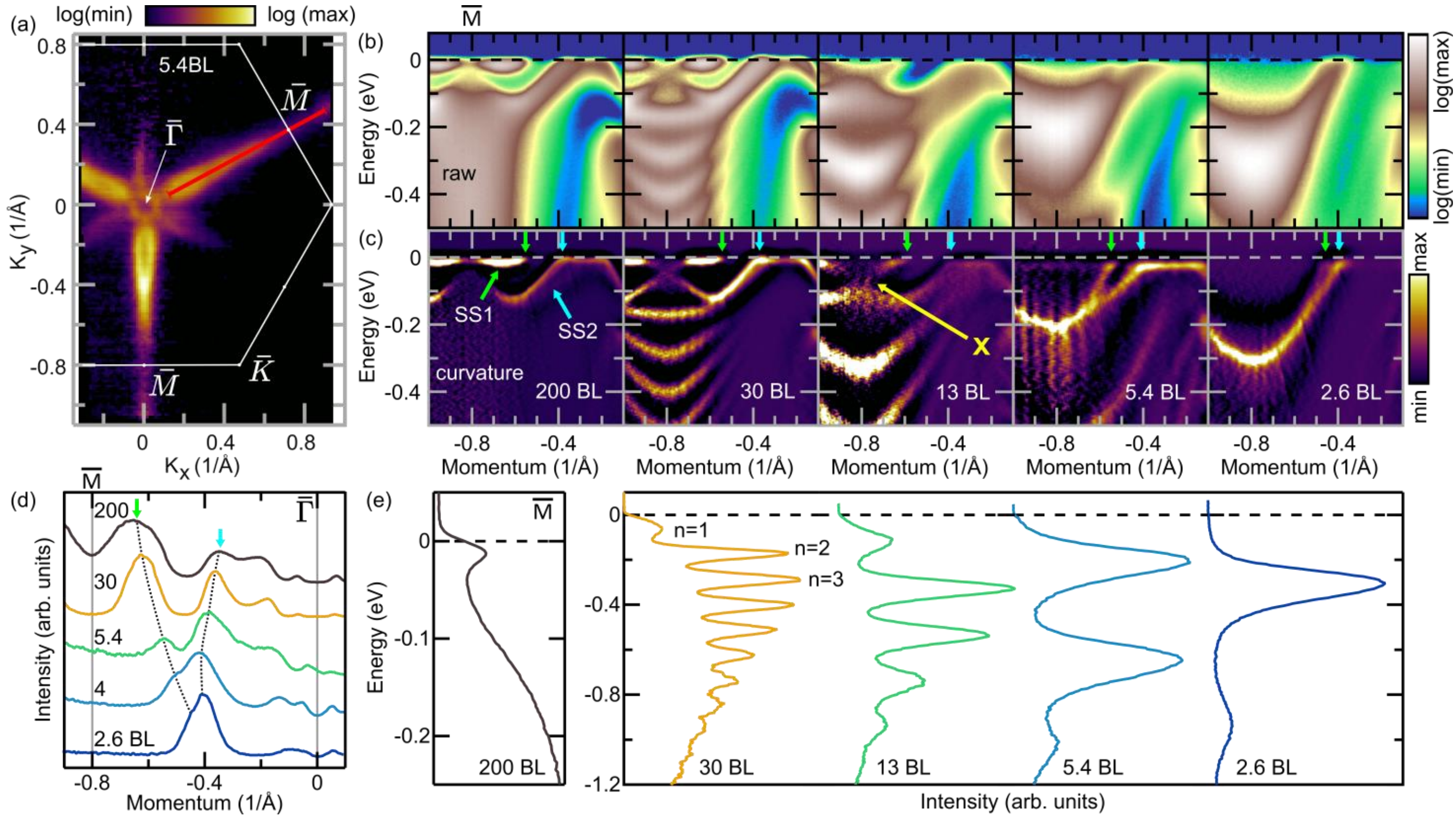

Fig. 3. Bi (111) surface states (SS1, SS2) and QWS measured with ARPES at $h\nu$ = 37.5 eV for various film thicknesses. (a) Fermi surface of a 5.4 BL thick Bi (111) film at $E_F$. The corresponding $E$–$k$ cut along the $\bar{\Gamma} - \bar{M}$ path in (b) and (c) is highlighted. (b) ARPES raw images and (c) curvature plots [31]. The observed surface state band degeneracy at $\bar{M}$ ($k = 0.8$ Å$^{-1}$), marked by X, is attributed to SIA. (d) Momentum distribution curves at $E_F$. The surface state Fermi level crossings near $\bar{M}$, highlighted by arrows in (c) and (d), decrease in separation as the films become thinner. (e) Energy distribution curves at $\bar{M}$ showing the energies of the QWS analyzed in Fig. 4(e).

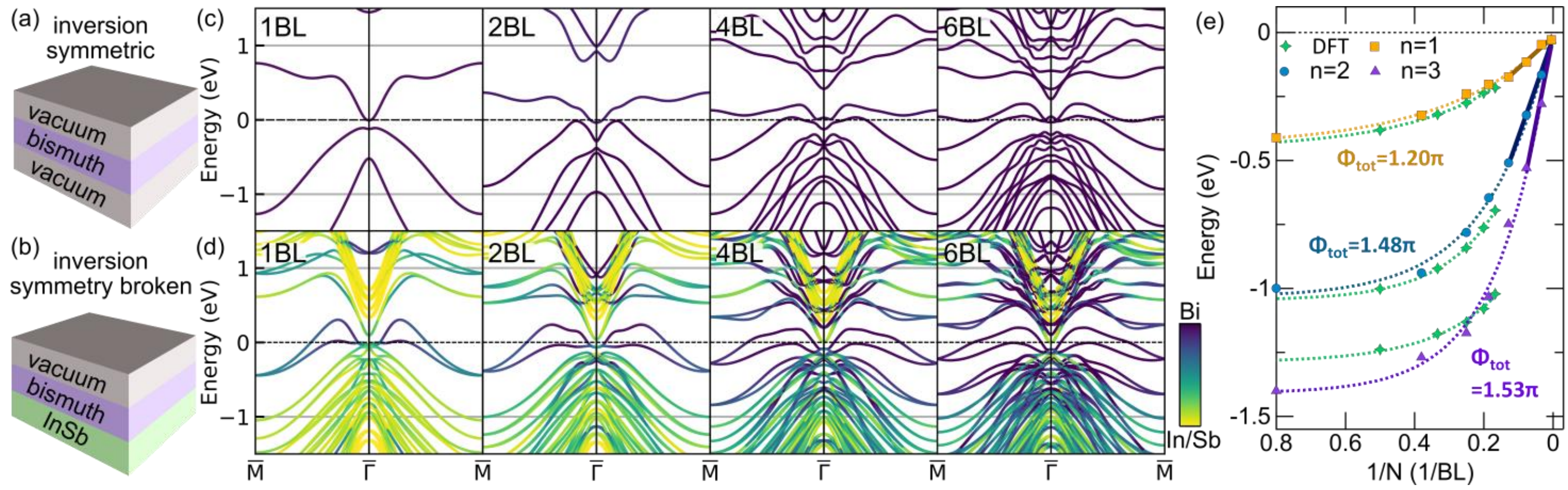

Fig. 4. Structure inversion symmetry in ultrathin Bi films. Schematics of stacks for (a) freestanding film preserving structure inversion symmetry and (b) film with broken inversion symmetry due to non-equivalent interfaces. DFT calculations for varying ultrathin film thicknesses for (c) freestanding Bi slab in (a), and (d) film with broken inversion symmetry in (b). In (d), all states with more than 60% of their spectral weight in the Bi film are marked in purple, and states with 0% of their weight in the Bi film (originating from the InSb layer) are marked in yellow. (e) DFT (green) and ARPES (yellow, blue, purple) QWS energy position at $\bar{M}$ vs inverse film thickness, 1/N. Full solid lines are linear fits used for analyzing the total phase shift (described in Section S3 [45]), and the dotted lines are drawn as guides for the eye.

Surface state crossing in thin films, which should be susceptible to surface state hybridization, can be explained by strong Bi-InSb interfacial bonding causing symmetry breaking. Note that Bi films grown on other substrates have not exhibited band degeneracy, which can be attributed to weak vdW-like interactions at the film-substrate interface [6]. Several experimental observations support strong Bi-InSb bonding. First, we note the epitaxial stabilization of ultrathin

(111) Bi on InSb for films as thin as 1 BL in Fig. 2(e), and the formation of a unique fractal structure that requires strong Bi-InSb bonding [48]. Moreover, the nucleation of tensile-strained and azimuthally aligned Bi films [49] to the underlying InSb substrate suggests that the bonding energy initially surpasses the elastic energy later gained when the film relaxes. Finally, ultraviolet photoemission measurements reveal a shift in both the In 4*d* and Sb 4*d* core levels upon Bi deposition in Fig. S4 [45], indicating the formation of Bi-Sb bonds. Therefore, we conclude that breaking the inversion symmetry by having strong bonding between the Bi film and the substrate leads to a decoupling of top and bottom surfaces and the degeneracy of the surface state at the $\bar{M}$ TRIM point, thus indicating a topologically trivial phase.

To further understand the nature of bonding between Bi and InSb, and whether remnant tensile biaxial strain could lead to a trivial gap at L [1] and influence the topological phase assignment, we examined the degree of in-plane relaxation predicted by DFT calculations (Fig. S7 and Section S3 [45]). The unstrained Bi/InSb structure reproduces the experimental trend of early film relaxation from 2 BL [Fig. 2(f)] [49] and shows no significant change in the surface state or QWS dispersions. Thus, we propose that the trivial topological assignment applies to Bi films grown on InSb with biaxial strain ranging from 0-0.8% tensile strain.

It is important to note that SIA not only controls surface state hybridization, but also can tune the 3D electronic structure, potentially altering the band order and destroying band inversion. Additional work is needed to determine how the topological order of Bi with SIA may differ from that of Bi in a 3D single crystal without SIA, and whether the starting substrate, InSb, could also alter the topological order of Bi. Band structure calculations for quantum wells that maintain the Bi thin film inversion symmetry should be compared to results in Fig. 4 for SIA structures. Experimental work with dual-gated Bi structures, as recently demonstrated for other topological systems [53], could also help control the displacement field across the thin films.

Another property affected by SIA and strain in ultrathin Bi films is the predicted bandgap opening in a 1 BL Bi film [28,54]. According to earlier predictions, tensile strained Bi films should undergo a semimetal to semiconducting transition (indirect gap *T-L*), resulting in a valence-band edge displaced to higher binding energies and a lowered surface state minima at $\bar{\Gamma}$ [1,23,25,28,32,52]. However, our Bi/InSb DFT calculations show that the surface state crossing at $\bar{\Gamma}$ moves closer to $E_F$ as the film thickness decreases [Fig. 4(d) and Section S2 [45]]. ARPES data in Fig. S5 [45] show that the surface state crossing energy at $\bar{\Gamma}$ approaches $E_F$ with decreasing film thickness, similar to ultrathin Sb/InSb(111)A films [55]. Film-substrate interactions at the few-BL limit could drive the valence band energy shift. Substrate-film interaction in Fig. 4(d) also results in trivial surface states intersecting $E_F$. Therefore, any 1D edge transport channel in the HOTI phase will coexist with trivial surface state conduction unless both interfaces are passivated. Our DFT calculations are consistent with ARPES measurements of a 1.3 BL thick film (Fig. S6 [45]), showing the surface states traversing $E_F$.

Next, we analyzed the QWS in Fig. 4(e) and Fig. S2 [45] using a phase accumulation model [3,8,32], as detailed in Section S3 [45]. At a thickness of 200 BL, QWS are not observed in Fig. 3(e) due to the small energy spacing, but thinner films display a larger QWS energy separation with decreasing film thickness. The binding energies at $\bar{M}$ of the top three quantum-well bands (*n*=1 being the surface state crossing point and *n*=2,3 the two following QWS) in the DFT calculations and ARPES measurements for varying Bi film thicknesses (N) are compared in Fig. 4(e), showing excellent agreement. For films thicker than 10 BLs, a linear-like region in Fig. 4(e) describes the expected $E$-$k_z$ bulk-like film dispersion [8]. Ultrathin Bi films <10 BL do not obey the linear relationship $E \propto \frac{1}{N}$, thus indicating a deviation from bulk-like dispersion along *X-L* [see Fig. 1(a)], suggesting a transition to 2D-like behavior. From the linear fits in Fig. 4(e), we extract the total phase shift, $\Phi_{tot}$: $\Phi_{tot}^{n=1} = 1.20\pi$, $\Phi_{tot}^{n=2} = 1.48\pi$, $\Phi_{tot}^{n=3} = 1.53\pi$ and the phase shift at the InSb-Bi interface ($\Phi_{Bi-InSb}$) at a 0.15 eV binding energy: $\Phi_{Bi-InSb}^{n=1} = 1.33\pi$, $\Phi_{Bi-InSb}^{n=2} = 1.61\pi$, $\Phi_{Bi-InSb}^{n=3} = 1.66\pi$. $\Phi_{tot}^{n=1}$ is close to the value reported for Bi films grown on Si [3], indicating a similar confining potential for the top surface state.

## 3. SUMMARY

In conclusion, we report the growth and evolution of surface state dispersion for large-area, single-domain oriented ultrathin films of Bi (111) synthesized on InSb (111)B. We find that strong film-substrate bonds stabilize ultrathin Bi films in the (111) orientation, offering a new route for the epitaxial growth and integration of other related topological systems, such as compressive strained Bi films and $Bi_{1-x}Sb_x$ 2D layers on insulating III-V substrates. We observe a surface state crossing at the $\bar{M}$ point in agreement with results of DFT calculations, suggesting a topologically trivial phase in the film. We also studied QWS through a phase accumulation model and showed a transition to 2D-like behavior for films <10 BL. Contrary to previous predictions of confinement [28] or strain-induced semimetal-to-semiconductor transition [56] in freestanding Bi thin film, we find that for the inversion-symmetry broken films, the surface states cross $E_F$ for all thicknesses down to 1 BL. For an unambiguous determination of the Bi film topological phase, additional studies will be necessary, targeting smoother films, additional film thicknesses, and incorporating state-of-the-art laser ARPES (with an energy resolution < 5 meV). Further analytical studies are also required, studying the role of SIA on surface state dispersion in 2D-like films of 3D TIs. Our work experimentally demonstrates the possibility of tailoring topological and trivial surface states in group-V 2D layers through heteroepitaxial interfaces. Despite numerous theoretical studies on the surface chemistry of buckled Bi films [40,41,57] and other elemental 2D materials [58], there are still few experimental reports on inorganic or molecular functionalization [59,60]. Future attempts to control surface terminations in 2D materials through overlayer growth could aid in band structure engineering of inversion-symmetric/asymmetric structures and in identifying topological phases and their transport signatures.

## ACKNOWLEDGMENTS

**Funding:** Initial growth studies, ARPES experiments, and theoretical work were supported by the U.S. Department of Energy (contract no. DE-SC0014388). This research used resources of the Advanced Light Source, which is a DOE Office of Science User Facility under contract no. DE-AC02-05CH11231. We acknowledge the use of shared facilities of the NSF Materials Research Science and Engineering Center (MRSEC) at the University of California Santa Barbara (NSF DMR-2308708). DFT calculations used the National Energy Research Scientific Computing Center (NERSC), a U.S. Department of Energy Office of Science User Facility operated under contract no. DE-AC02-05CH11231. H. S. I. gratefully acknowledges support from the UC Santa Barbara NSF Quantum Foundry funded via the Q-AMASE-i program under award DMR-1906325 and support for further growth studies and development of the vacuum suitcases. M.Z. And A.J. acknowledge funding from NSF Award #OIA- 2217786 and NSF UDCHARM University of Delaware Materials Research Science and Engineering Center (MRSEC) Grant No. DMR2011824. S.K. acknowledges funding from LDRD under U.S. Department of Energy Prime Contract No. DE-AC02-09CH11466. **Author contributions:** H.S.I. and C.J.P. conceived the study. H.S.I. performed thin film growth, ARPES measurements, and data analysis. J.T.D. performed STM measurements with assistance from C.P.D. . A.N.E., Y.C., S.N., and A.V.F. assisted in ARPES experiments. DFT calculations were conducted by M.Z. under the supervision of S.K. and A.J. The authors thank Dai Q. Ho for useful discussions. A.N.E. and C.P.D. designed ultra-high vacuum components and sample holders. The manuscript was prepared by H.S.I. All authors discussed the results and commented on the manuscript. **Competing interests:** The authors declare that they have no competing interests. **Data and materials availability:** All data needed to evaluate the conclusions in the paper are present in the paper and/or the Supplementary Materials. Additional data related to this paper may be requested from the authors.